\documentclass[12pt]{elsarticle}

\usepackage{amssymb}
\usepackage{graphicx}
\usepackage{amsmath}
\usepackage{subfigure}
\usepackage{color}
\usepackage{multirow}
\biboptions{sort&compress}
\journal{Journal of magnetism and magnetic materials}

\begin{document}

\begin{frontmatter}



\title{Gapped paramagnetic state in a frustrated spin-$\frac{1}{2}$ Heisenberg antiferromagnet on the cross-striped square lattice}


\author{P H Y Li$^{1,2}$ and R F Bishop$^{1,2}$}
\ead{peggyhyli@gmail.com; raymond.bishop@manchester.ac.uk}

\address{$^{1}$ School of Physics and Astronomy, Schuster Building, The University of Manchester, Manchester, M13 9PL, UK}

\address{$^{2}$ School of Physics and Astronomy, University of Minnesota, 116 Church Street SE, Minneapolis, Minnesota 55455, USA}

\begin{abstract}
 We implement the coupled cluster method to very high orders
  of approximation to study the spin-$\frac{1}{2}$ $J_{1}$--$J_{2}$
  Heisenberg model on a cross-striped square lattice.  Every
  nearest-neighbour pair of sites on the square lattice has an
  isotropic antiferromagnetic exchange bond of strength $J_{1}>0$,
  while the basic square plaquettes in alternate columns have either
  both or neither next-nearest-neighbour (diagonal) pairs of sites
  connected by an equivalent frustrating bond of strength
  $J_{2} \equiv \alpha J_{1} > 0$.  By studying the magnetic order
  parameter (i.e., the average local on-site magnetization) in the
  range $0 \leq \alpha \leq 1$ of the frustration parameter we find
  that the quasiclassical antiferromagnetic N\'{e}el and (so-called)
  double N\'{e}el states form the stable ground-state phases in the
  respective regions $\alpha < \alpha_{1a}^{c} = 0.46(1)$ and
  $\alpha > \alpha_{1b}^{c} = 0.615(5)$.  The double N\'{e}el state
  has N\'{e}el ($\cdots\uparrow\downarrow\uparrow\downarrow\cdots$)
  ordering along the (column) direction parallel to the stripes of
  squares with both or no $J_{2}$ bonds, and spins alternating in a
  pairwise
  ($\cdots\uparrow\uparrow\downarrow\downarrow\uparrow\uparrow\downarrow\downarrow\cdots$)
  fashion along the perpendicular (row) direction, so that the
  parallel pairs occur on squares with both $J_{2}$ bonds present.
  Further explicit calculations of both the triplet spin gap and the
  zero-field uniform transverse magnetic susceptibility provide
  compelling evidence that the ground-state phase over all or most of
  the intermediate regime $\alpha_{1a}^{c} < \alpha < \alpha_{1b}^{c}$
  is a gapped state with no discernible long-range magnetic order. 
\end{abstract}


\end{frontmatter}


\section{Introduction}
\label{introd_sec}
Frustrated quantum magnets, in which competing types of interactions
vie with one another, even at the classical level, to promote
different forms of magnetic long-range order (LRO), provide
notoriously difficult challenges for any computational many-body
technique \cite{Scholl:2004_2Dmagnetism}.  At the theoretical level the underlying spin-lattice
models are of huge interest since their zero-temperature ($T=0$) phase
diagrams (in the space spanned by the parameters describing the
interaction strengths of the competing interactions in the model
Hamiltonian) often display a rich variety of phases with orderings
that have no classical counterparts.  The resulting quantum phase
transitions provide a challenging theoretical and computational arena
in which to describe accurately the physical properties of the models
near their corresponding quantum critical points (QCPs) at which
long-range many-body quantum entanglement, driven by frustration, is
paramount \cite{Sachdev:2011_QPT}.  At the same time such spin-lattice models are also of
great interest experimentally, since they often mimic rather
accurately the properties of a large variety of real materials.

Of particular interest in this respect are two-dimensional (2D)
models.  This is partly because the effects of quantum fluctuations
are higher in systems with reduced dimensionality, and also because in
recent years a large number of layered quasi-2D real magnetic
materials have been synthesised in which the strengths of the
intralayer magnetic interactions are much larger than those acting
between layers.  Furthermore, many such synthetic 2D spin-lattice
systems are now increasingly being modelled rather accurately using
the properties of suitable ultracold atoms trapped in 2D optical
lattices.  The great advantage of the latter is that in such
experiments one can also tune the system so as effectively to vary the
relative strengths of the competing interaction terms in the
Hamiltonian.  In this way one may attempt to sweep over any pertinent
QCPs \cite{Sachdev:2008_QPT}.  One role of the theorist in this endeavour is then to suggest
suitably interesting models for the experimentalist to attempt to
mimic.

One particularly interesting, and especially challenging, model in
this respect is the $J_{1}$--$J_{2}$ model on the 2D square lattice in
which antiferromagnetic (AFM) nearest-neighbour (NN) isotropic ($XXX$)
Heisenberg interactions with exchange coupling strength parameter
$J_{1}>0$ compete with next-nearest-neighbour (NNN) $XXX$ Heisenberg
interactions with respective strength parameter $J_{2}>0$.  This
archetypal model, which was introduced nearly 30 years ago in an
attempt to describe the disappearance of AFM N\'{e}el LRO in the
cuprate high-temperature superconductors
\cite{Inui:1988_cuprate_highT_supercondctr,Chandra:1988,Dagotto:1989},
has since then received huge theoretical interest.  In particular,
many different theoretical and computational techniques have been used
to study it (see, e.g.,
Refs.\ \cite{Schulz:1992,Schulz:1996,Richter:1993,Richter:1994,Zhitomirsky:1996,Bishop:1998_J1J2mod,Singh:1999,Capriotti:2001,Sirker:2006,Schm:2006_stackSqLatt,Mambrini:2006,Bi:2008_JPCM_J1J1primeJ2,Bi:2008_PRB_J1xxzJ2xxz,Darradi:2008_J1J2mod,Murg:2009_peps,Richter:2010_ED40,Reuther:2010_J1J2mod,Yu:2012,Jiang:2012,Mezzacapo:2012,LiT:2012,Wang:2013,Hu:2013,Zhang:2013:J1J2SqLatt,Gong:2014_J1J2mod_sqLatt,Doretto:2014_J1J2mod_sqLatt,Qi:2014_J1J2SqLatt,Metavitsiadis_Eggert:2014_J1J2mod_sqLatt,Ren:2014_J1J2SqLatt,Wing:2014_J1J2SqLatt,Chou:2014_J1J2SqLatt,Morita:2015_J1J2SqLatt,Richter:2015_ccm_J1J2sq_spinGap}).
These include several very successful applications
\cite{Bishop:1998_J1J2mod,Schm:2006_stackSqLatt,Bi:2008_JPCM_J1J1primeJ2,Bi:2008_PRB_J1xxzJ2xxz,Darradi:2008_J1J2mod,Richter:2015_ccm_J1J2sq_spinGap}
of the coupled cluster method (CCM).

Of comparable theoretical interest and equally challenging, but less
studied, is the family of half-depleted square-lattice
$J_{1}$--$J_{2}$ models, in which half of the NNN $J_{2}$ bonds are
removed in an ordered fashion.  Two of the most interesting members of
this class are the so-called anisotropic planar pyrochlore (or
anisotropic checkerboard model) and the cross-striped square-lattice
model, in both of which half of the basic square plaquettes formed
from four NN $J_{1}$ bonds have both diagonal $J_{2}$ bonds retained
(i.e., the filled squares) while the rest have neither (i.e., the
empty squares).  They differ in that, while in the former model the
empty and filled squares alternate along both rows and columns of the
square lattice, the latter comprises alternating columns (or,
equivalently, rows) of filled and empty squares.  While the
anisotropic planar pyrochlore has been studied by a variety of methods
in the past including the CCM (see, e.g., Refs.\
\cite{Bishop:2012_checkerboard,Li:2015_checkboard_fullPhase} and
references contained therein), the equally interesting cross-striped
square-lattice model has, somewhat surprisingly, received only scant
attention \cite{Bishop:2013_crossStripe}.  As a consequence we apply
the CCM to it here in order to investigate further its $T=0$ phase
diagram.  Our main finding will be to confirm that the
spin-$\frac{1}{2}$ model has an intermediate paramagnetic state with
no obvious magnetic LRO that lies between the two classical states
with AFM order in the frustrated case where $J_{1}>0$ and $J_{2}>0$.
Furthermore, we find compelling evidence that this intermediate state
is gapped.

\section{The model}
\label{model_sec}
\begin{figure}[t]
\begin{center}
\vspace{0.7cm}
\hspace{-0.3cm}
\mbox{
\subfigure[]{\scalebox{0.4}{\includegraphics{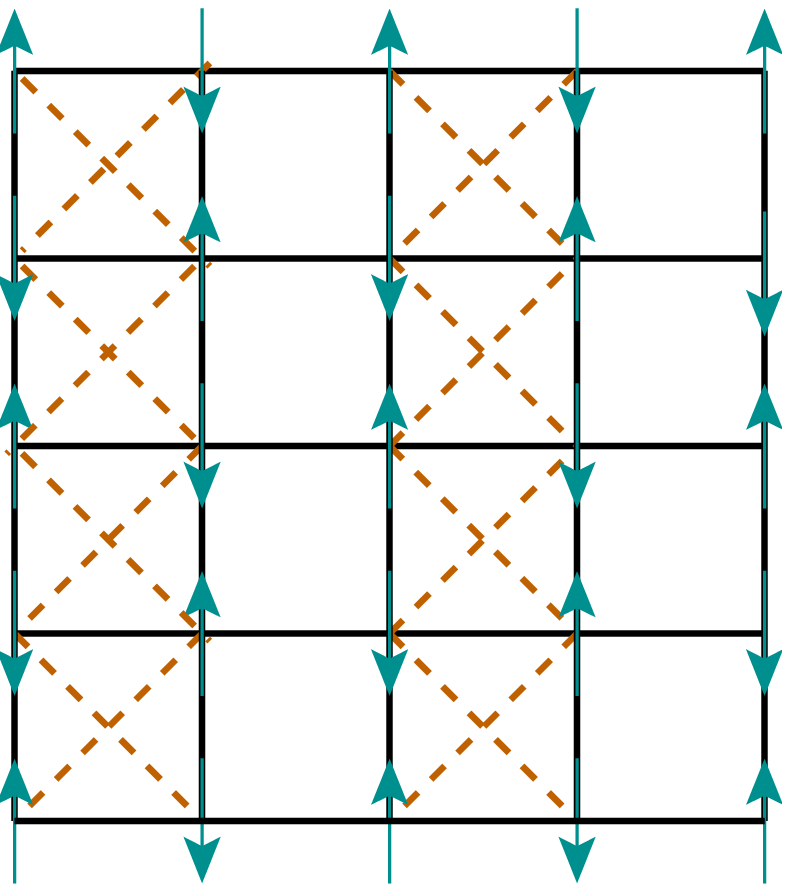}}} \hspace{0.05cm}
\subfigure[]{\scalebox{0.4}{\includegraphics{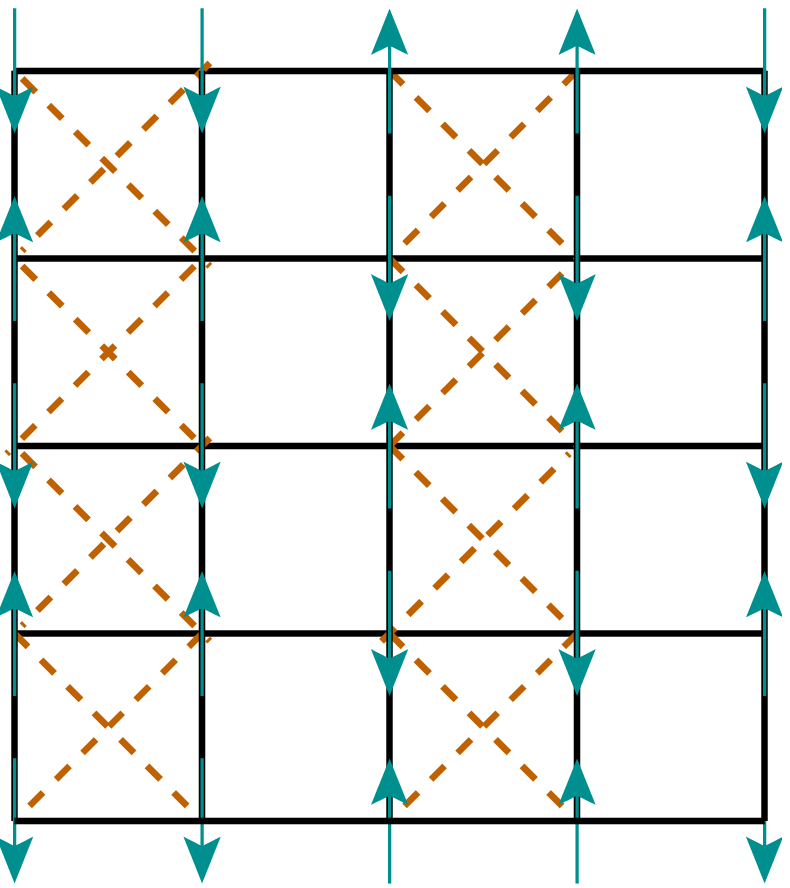}}}  \hspace{0.05cm}
}
\end{center}
\caption{The $J_{1}$--$J_{2}$ Heisenberg model on a cross-striped
  square lattice.  The solid (black) lines are $J_{1}$ bonds and the
  dashed (brown) lines are $J_{2}$ bonds.  The (cyan) arrows represent the
  relative spin directions in: (a) the N\'{e}el state, and (b) the double
  N\'{e}el (DN) state.}
\label{model_bonds}
\end{figure}

We study the Hamiltonian 
%%%%%%%%%%%%%%%%%%%%
\begin{equation}
H = J_{1}\sum_{\langle i,j \rangle} \mathbf{s}_{i}\cdot\mathbf{s}_{j} + J_{2}\sum_{\langle\langle i,k \rangle\rangle'} 
\mathbf{s}_{i}\cdot\mathbf{s}_{k}\,, \label{H_eq}
\end{equation}
%%%%%%%%%%%%%%%
where the operators
$\mathbf{s}_{i}\equiv(s^{x}_{i},s^{y}_{i},s^{z}_{i}$) are spin-$s$
SU(2) operators on each of the $N \rightarrow \infty$ sites $i$ of a
2D square lattice.  We consider here the extreme quantum case
$s=\frac{1}{2}$.  The sum over $\langle i,j \rangle$ and
$\langle\langle i,k \rangle\rangle'$ run respectively over all
distinct NN bonds and over half of the distinct NNN (diagonal) bonds
in the cross-striped pattern shown in Fig.\ \ref{model_bonds}.
We are interested here in investigating the
$T=0$ phase diagram of the spin-$\frac{1}{2}$ model when both bonds
are AFM in nature (i.e., $J_{1}>0$, $J_{2} \equiv \alpha J_{1}>0$),
and hence act to frustrate one another.

The phase diagram for the model in the classical
($s \rightarrow \infty$) limit has been discussed in detail in Ref.\ 
\cite{Bishop:2013_crossStripe}.  In our region of interest there are
two stable AFM ground-state (GS) phases, shown in Figs.\
\ref{model_bonds}(a) and \ref{model_bonds}(b) respectively, separated
by a first-order transition at the critical point
$\alpha^{{\rm cl}}=\frac{1}{2}$.  For values
$\alpha < \alpha^{{\rm cl}}$ of the frustration parameter the
classical system has N\'{e}el LRO, as shown in Fig.\
\ref{model_bonds}(a).  However, for values
$\alpha > \alpha^{{\rm cl}}$, another collinear AFM state, the
so-called double N\'{e}el (DN) state shown in Fig.\
\ref{model_bonds}(b), forms the stable GS phase.  It has AFM N\'{e}el
ordering along the direction parallel to the stripes of filled squares
(viz., along columns in Fig.\ \ref{model_bonds}), together with spins
alternating in a pairwise fashion along the perpendicular direction.
Thus NN spins are parallel on filled squares and antiparallel on empty
squares in the row direction in Fig.\ \ref{model_bonds}(b).

In the present study we will employ both of these classical GS AFM
phases as reference states.  Neither of them are eigenstates of the
quantum Hamiltonian for the case $s=\frac{1}{2}$ under study, but we will use the CCM
as a systematic tool to incorporate the quantum many-body correlations
on top of them within a controlled approximation hierarchy that
asymptotically becomes exact as one approaches the infinite-order
limit.  We will implement the method computationally to high orders of
approximation and then use well-known extrapolation schemes to
calculate various physical quantities for the model, including the
magnetic order parameter, the zero-field transverse magnetic
susceptibility and the triplet spin gap.

In order to calculate the susceptibility we place the system in a
transverse uniform magnetic field ${\mathbf h} = h\hat{x}_{s}$, where
$\hat{x}_{s}$ is a unit vector in the spin-space $x$-direction.  The
Hamiltonian $H \equiv H(h=0)$ of Eq.\ (\ref{H_eq}) then becomes
$H(h) = H(0) - h\sum_{k=1}^{N}s_{k}^{x}$, in units where the
gyromagnetic ratio $g\mu_{B}/\hbar=1$.  The spins, previously aligned
as shown in Figs.\ \ref{model_bonds}(a) and \ref{model_bonds}(b),
respectively, in the N\'{e}el and DN states, now cant at an angle
$\theta = \theta(h)$ with respect to their zero-field configurations
along the spin-space $z$-direction.  These canted states are used as
our CCM reference states in the case of an applied external magnetic
field, and $\theta$ is calculated in practice
\cite{Farnell:2009_Xcpty_ExtMagField} so as to minimize the energy
expectation value $E(h)$ for a given value of $h$ at each level of CCM
approximation discussed below.  We define the (uniform) transverse
magnetic susceptibility as usual by
$\chi(h)=-N^{-1}{\rm d}^{2}E/{\rm d}h^{2}$, and its zero-field limit
as $\chi \equiv \chi(0)$.  It is simple to calculate $\theta(h)$ and
hence $\chi$ for the two classical $(s \rightarrow \infty)$ AFM states
in Fig.\ \ref{model_bonds}, thereby obtaining the classical result for
the model,
\begin{equation}
J_{1}\chi^{{\rm cl}} = \left \{ \begin{array}{ll}
\frac{1}{8}; &  \alpha \leq \frac{1}{2}\,, \\ [0.3em]
\frac{1}{2}(3+2\alpha)^{-1}; & \alpha \geq \frac{1}{2}\,,
\end{array}
\right.          \label{ExtMField_classical}
\end{equation}
where $\alpha \equiv J_{2}/J_{1}$ is the frustration parameter, and $J_{1}>0$.

\section{The coupled cluster method}
\label{ccm_sec}
The CCM
\cite{Bishop:1991_TheorChimActa_QMBT,Bishop:1998_QMBT_coll,Zeng:1998_SqLatt_TrianLatt,Fa:2004_QM-coll}
is one of the most flexible and most accurate techniques of
microscopic quantum many-body theory.  It is both size-consistent and
size-extensive, and thereby provides results in the infinite-lattice
($N \rightarrow \infty$) limit at every level of approximation.  The
quantum correlations are expressed in a suitably exponentiated form,
which is the hallmark of the CCM, on top of suitably chosen model (or
reference) states (and see, e.g., Refs.\
\cite{Bishop:1998_J1J2mod,Zeng:1998_SqLatt_TrianLatt,Fa:2004_QM-coll,Bishop:2000_XXZ}).
We use here the N\'{e}el and DN states shown in Fig.\
\ref{model_bonds}, together with their canted versions for the
calculation of $\chi$.  For the current $s=\frac{1}{2}$ model we also
use the extensively used and very well-tested localized
lattice-animal-based subsystem (LSUB$m$) truncation hierarchy of
approximation
\cite{Zeng:1998_SqLatt_TrianLatt,Fa:2004_QM-coll,Bishop:2000_XXZ}.  At
the corresponding $m$th level of approximation one retains all
multispin correlations in the CCM correlation operators that are
defined over all distinct locales (or lattice animals) on the lattice
that comprise $m$ or fewer contiguous sites (i.e., those in which
every site is NN to at least one other).  Very importantly, the method
preserves both the Goldstone linked-cluster theorem and the
Hellmann-Feynman theorem at every LSUB$m$ level of approximation.  The
techniques involved in deriving and solving the large and inherently
nonlinear sets of coupled CCM equations at high LSUB$m$ orders have
been extensively discussed in the literature (see, eg., Refs.\
\cite{Bishop:1998_J1J2mod,Zeng:1998_SqLatt_TrianLatt,Fa:2004_QM-coll,Bishop:2000_XXZ}),
and will not be repeated here.  

Space- and point-group symmetries of both the Hamiltonian and the CCM
reference state, together with any conservation laws, are used to
reduce the set of independent spin configurations retained in the CCM
LSUB$m$ correlation operators to a minimal number, $N_{f}(m)$.
Nevertheless, $N_{f}$ increases rapidly as the truncation index $m$ is
increased.  Accordingly, one needs both considerable supercomputing
resources and massive parallelization for the higher-order
calculations \cite{Zeng:1998_SqLatt_TrianLatt,ccm_code}.  For the
present spin-$\frac{1}{2}$ model we are able to perform LSUB$m$
calculations for $m \leq 10$ for both the ground and lowest-lying
spin-triplet excited states based on either the N\'{e}el or DN AFM
states as CCM model states, and using the cross-striped square-lattice
geometry in which all pairs of sites connected by either $J_{1}$ or
$J_{2}$ bonds are considered as NN pairs for the purposes of
contiguity contained in the selection of LSUB$m$ multispin clusters.
For example, for the GS results (in the absence of any external
magnetic field) at the LSUB10 level, we have
$N_{f}=768\,250$ $(853\,453)$ using the N\'{e}el (DN) reference states.
Similarly, for the LSUB10 calculation of the triplet spin gap
$\Delta$, we have $N_{f}=1\,400\,161$ ($1\,500\,377$) using the
N\'{e}el (DN) reference states.  Due to the much reduced symmetries of
the respective canted states in the presence of an external magnetic
field, we are only able to perform LSUB$m$ calculations for $\chi$
with $m \leq 8$.  Thus, we have $N_{f}=203\,508$ ($203\,445$) at the
LSUB8 level for calculations based on the canted N\'{e}el (DN)
reference states.

The {\it only} approximation made in our CCM calculations comes in the
final step of extrapolation to the exact $m \rightarrow \infty$ limit.
Although no exact such schemes are known, by now a great deal of
experience has been accumulated from an extremely large number of
successful applications of the CCM to many spin-lattice models (see,
e.g.,
\cite{Schm:2006_stackSqLatt,Bi:2008_JPCM_J1J1primeJ2,Bi:2008_PRB_J1xxzJ2xxz,Darradi:2008_J1J2mod,Richter:2015_ccm_J1J2sq_spinGap,Bishop:2012_checkerboard,Li:2015_checkboard_fullPhase,Bishop:2013_crossStripe,Farnell:2009_Xcpty_ExtMagField,Zeng:1998_SqLatt_TrianLatt,Fa:2004_QM-coll,Bishop:2000_XXZ}
and references contained therein).  Thus, for example, for the LSUB$m$
approximation $M(m)$ of the magnetic order parameter (viz., the
average local on-site magnetization) $M$ we use the scheme
\begin{equation}
M(m) = \mu_{0}+\mu_{1}m^{-1/2}+\mu_{2}m^{-3/2}\,,   \label{M_extrapo_frustrated}
\end{equation}
which has been found to be appropriate for highly frustrated systems,
particularly in the vicinity of a QCP, to extract the extrapolated
value $\mu_{0}$ for $M$.  Similarly, very well-tested schemes for the
LSUB$m$ approximants $\Delta(m)$ and $\chi(m)$ of the respective triplet spin gap
and the zero-field transverse magnetic susceptibility $\chi$ are
\begin{equation}
\Delta(m) = d_{0}+d_{1}m^{-1}+d_{2}m^{-2}\,,   \label{Eq_spin_gap}
\end{equation}
and
\begin{equation}
\chi(m) = x_{0}+x_{1}m^{-1}+x_{2}m^{-2}\,,   \label{Eq_X}
\end{equation}
and we utilize both schemes here to extract the extrapolated values
$d_{0}$ and $x_{0}$, respectively.

\section{Results and Discussion}
\label{results_sec}
\begin{figure}
\begin{center}
\includegraphics[width=8.5cm]{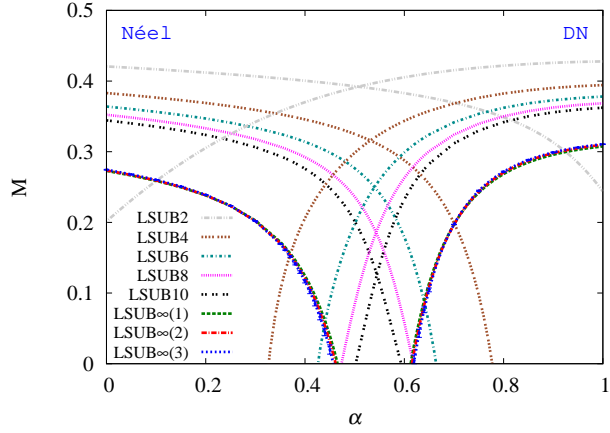}
\caption{CCM results for the GS magnetic
  order parameter $M$ as a function of the frustration parameter
  $\alpha \equiv J_{2}/J_{1}$ for the spin-$\frac{1}{2}$ $J_{1}$--$J_{2}$ model
  on a cross-striped square lattice, with $J_{1}>0$.  Results are shown based on both the quasiclassical N\'{e}el
  (left curves) and DN (right curves) AFM states used as the CCM model state, in LSUB$m$ approximations with $m=2,4,6,8,10$.  Also shown are two corresponding sets of three LSUB$\infty(i)$ extrapolations, each based on Eq.\ (\ref{M_extrapo_frustrated}) but using as input the separate respective data sets with $m=\{2,6,10\}$ for $i=1$, $m=\{4,6,8\}$ for $i=2$ and $m=\{4,6,8,10\}$ for $i=3$.}
\label{M}
\end{center}
\end{figure}

We first present in Fig.\ \ref{M} our CCM results for the GS magnetic
order parameter $M$ of the spin-$\frac{1}{2}$ $J_{1}$--$J_{2}$
Heisenberg antiferromagnet on a cross-striped square lattice.  
Results at LSUB$m$ levels of approximation with $m \leq 10$ are shown,
based on both the quasiclassical N\'{e}el and DN states illustrated in
Figs.\ \ref{model_bonds}(a) and \ref{model_bonds}(b), respectively,
used as CCM model states.  We also display in Fig.\ \ref{M} the
corresponding ($m \rightarrow \infty$) LSUB$\infty$ extrapolations
obtained from the use of Eq.\ (\ref{M_extrapo_frustrated}).  In order
to illustrate the robustness of the extrapolation, which we reiterate
is the {\it sole} approximation that we make, we show explicitly for both
AFM phases three separate such extrapolations, based on the respective
LSUB$m$ data sets $m=\{2,6,10\}$, $m=\{4,6,8\}$ and $m=\{4,6,8,10\}$,
each used separately as input to Eq.\ (\ref{M_extrapo_frustrated}).

It is clear from Fig.\ \ref{M} that both quasiclassical GS phases lose
their stability in a regime around the point
$\alpha^{\rm cl}=\frac{1}{2}$ where the classical
($s\rightarrow \infty$) version of the model shows a phase transition
between the two AFM phases.  Indeed, the extrapolations show clear and
consistent evidence for a phase in the intermediate range
$\alpha^{c}_{1a} < \alpha < \alpha^{c}_{1b}$ with no classical
counterpart.  For $\alpha < \alpha^{c}_{1a}$ the
system has a GS phase with N\'{e}el AFM LRO, whereas for
$\alpha > \alpha^{c}_{1b}$ the GS phase has DN AFM
LRO, with $\alpha^{c}_{1a} < \alpha^{{\rm cl}} < \alpha^{c}_{1b}$.

In order to fit extrapolation schemes of the forms of Eqs.\
(\ref{M_extrapo_frustrated})--(\ref{Eq_X}) that contain three fitting
parameters it is clearly preferable to use input data sets comprising
greater than three values.  In this way we can also calculate the
error bars associated with the least-squares fits.  These are shown,
for example, in the LSUB$\infty(3)$ extrapolation shown in Fig.\
\ref{M}, which is based on the LSUB$m$ input data set with
$m=\{4,6,8,10\}$ for the calculated values of $M(m)$ at each value of
$\alpha$.  The errors associated with the fit are generally extremely
small, indicating its overall accuracy, although they do become
somewhat larger in the vicinity of the QCPs at $\alpha^{c}_{1a}$ and
$\alpha^{c}_{1b}$ where $M$ becomes small.  The numerical values of
the two phase transition points obtained from the LSUB$\infty(3)$
extrapolation are $\alpha^{c}_{1a}=0.457$ and $\alpha^{c}_{1b}=0.619$.

As a further demonstration of the extremely robust nature of the
extrapolation procedure we also show in Fig.\ \ref{M} two other fits
using only three input data points, namely LSUB$\infty(1)$ using the
LSUB$m$ input data set with $m=\{2,6,10\}$ and LSUB$\infty(2)$ with
$m=\{4,6,8\}$.  Both are in very close agreement with the
intrinsically more accurate LSUB$\infty(3)$ fit that uses the four
input data points $m=\{4,6,8,10\}$, even the LSUB$\infty(1)$ fit that
includes the lowest-order LSUB$m$ approximant with $m=2$.  The
corresponding values obtained for the two phase transition points are
$\alpha^{c}_{1a}=0.466$ and $\alpha^{c}_{1b}=0.613$ from the
LSUB$\infty(1)$ fit, and $\alpha^{c}_{1a}=0.462$ and
$\alpha^{c}_{1b}=0.616$ from the LSUB$\infty(2)$ fit.  A more detailed
sensitivity analysis using these (and other similar such) fits yields
our best values, $\alpha^{c}_{1a}=0.46(1)$ and
$\alpha^{c}_{1b}=0.615(5)$, as obtained from our CCM calculations for
the vanishing of the magnetic order parameter $M$ in the N\'{e}el and
DN AFM phases, respectively.

The question that now immediately arises is what is the nature of the
intermediate phase without magnetic LRO?  Perhaps the simplest but
certainly not the only possible scenario in this case for the loss of
LRO is that the spin-spin correlation function between two spins at
lattice sites $k$ and $l$ decays exponentially as a function of the
distance $|{\mathbf r}_{k}-{\mathbf r}_{l}|$ between them on the
lattice,
\begin{equation}
|\langle {\mathbf s}_{k}\cdot{\mathbf s}_{l} \rangle| \propto {\rm exp}(-i|{\mathbf r}_{k}-{\mathbf r}_{l}|/\xi)\,.
\end{equation}
A system in such a phase with a finite correlation length $\xi$
behaves just as a finite system of comparable linear size, and hence
the quantized energy levels are discrete.  Accordingly, such a system
will have no low-energy excitations.  Its ground state will be a spin
singlet, just as we have assumed in our CCM calculations for the two
quasiclassical AFM states, and the lowest-lying triplet excited state
will lie at a finite energy, $\Delta > 0$, above the ground state.  In
order to calculate the triplet spin gap $\Delta$ we now perform
excited-state CCM calculations based on spin-triplet model states
obtained from the two quasiclassical AFM states (i.e., the N\'{e}el
and DN states) by flipping a single spin-$\frac{1}{2}$ particle.

\begin{figure}
\begin{center}
\includegraphics[width=8.5cm]{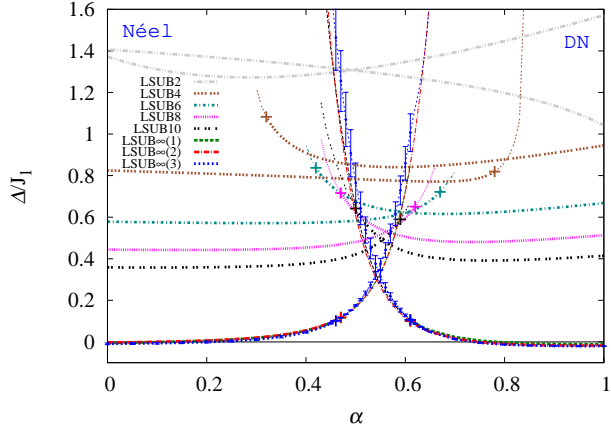}
\caption{CCM results for the triplet spin gap $\Delta$ (in units of $J_{1}$) as a function of the frustration parameter
  $\alpha \equiv J_{2}/J_{1}$ for the spin-$\frac{1}{2}$ $J_{1}$--$J_{2}$ model
  on a cross-striped square lattice, with $J_{1}>0$.  Results are shown based on both the quasiclassical N\'{e}el
  (left curves) and DN (right curves) AFM states with a single spin-flip used as the CCM model state, in LSUB$m$ approximations with $m=2,4,6,8,10$.  Also shown are two corresponding sets of three LSUB$\infty(i)$ extrapolations, each based on Eq.\ (\ref{Eq_spin_gap}) but using as input the separate respective data sets with $m=\{2,6,10\}$ for $i=1$, $m=\{4,6,8\}$ for $i=2$ and $m=\{4,6,8,10\}$ for $i=3$.  The points where the respective CCM solution has $M \rightarrow 0$ are shown by plus ($+$) symbols, and those unphysical portions of the curves beyond these points, where $M < 0$, are indicated by thinner lines.}
\label{E_gap}
\end{center}
\end{figure}

We show in Fig.\ \ref{E_gap} our LSUB$m$ results for the triplet spin
gap $\Delta$ with $m \leq 10$ based on both the N\'{e}el and DN
quasiclassical states with a single spin-flip, used separately as CCM
model states.
We also display three LSUB$\infty(i)$ extrapolations, each using the
fitting scheme of Eq.\ (\ref{Eq_spin_gap}) and with the same three
different input data sets as used in Fig.\ \ref{M} for $M$.  For the
N\'{e}el curves we first solve the CCM equations for the unfrustrated
case $\alpha=0$ to find the stable physical solution at each LSUB$m$
order.  Each such LSUB$m$ solution is then tracked, as the frustration
parameter $\alpha$ is incremented in small steps, out to the point
where the solution terminates.  The termination points for the
excited-state equations can be seen from Fig.\ \ref{E_gap} to decrease
monotonically as the termination order $m$ increases.  They are just
the analogues of those also typically found in the GS CCM equations.
They are simply manifestations of the corresponding QCP in the system
at which magnetic LRO, of the form implicit in the particular model
state being used, melts, as have been seen and well discussed in many
previous applications (see, e.g., Refs.\
\cite{Bi:2008_PRB_J1xxzJ2xxz,Bishop:2012_checkerboard,Li:2015_checkboard_fullPhase,Bishop:2013_crossStripe,Fa:2004_QM-coll}
and references cited therein).  As is commonly seen, each LSUB$m$
N\'{e}el solution terminates at a greater value of $\alpha$ than the
exact (LSUB$\infty$) value $\alpha^{c}_{1a}$.  Hence, we are able to
consider values for $\alpha$ that are appreciably beyond the actual
transition at $\alpha^{c}_{1a}$ into the quantum paramagnetic phase.
A similar situation also occurs on the other side of the transition,
where we now track each LSUB$m$ solution based on the DN model state
from high values of $\alpha$ down to a lower termination point.
Again, as seen from Fig.\ \ref{E_gap}, we are able to enter the region
below the corresponding QCP at $\alpha^{c}_{1b}$ for all finite values
of the LSUB$m$ truncation index $m$.

Clearly, parts of those regions beyond the respective QCPs that we may
enter with our LSUB$m$ solutions based on particular model states are
likely to be unphysical in the sense that the corresponding values for
the order parameter $M$ become negative.  For both the individual
LSUB$m$ solutions and the LSUB$\infty$ extrapolations we mark on the
curves by plus ($+$) symbols the points where $M \rightarrow 0$, as
determined from the corresponding curves in Fig.\ \ref{M}.  The
unphysical regions of the curves beyond these points where $M < 0$ are
shown in Fig.\ \ref{E_gap} by thinner curves than the respective
physical regions where $M > 0$, which are shown as thicker.

By comparing the three different LSUB$\infty(i)$ curves shown in Fig.\
\ref{E_gap} we see once again how extremely robust is our
extrapolation procedure for $\Delta$.  The least-squares-fit error
bars shown on the LSUB$\infty(3)$ extrapolation are also again very
small, except deep in the unphysical ($M < 0$) region for both the
N\'{e}el and DN curves beyond their totally unphysical crossing point.
The extrapolated results clearly show that for values of $\alpha$
appreciably below $\alpha^{c}_{1a}$ on the N\'{e}el side and above
$\alpha^{c}_{1b}$ on the DN side, the value of $\Delta$ is zero within
extremely small errors.  This is a very good independent check on our
extrapolation procedure since in both regimes we have magnetic LRO,
and the consequent low-energy magnon states (i.e., the soft Goldstone
modes) are gapless.  By contrast, in the intermediate regime, the GS
phase clearly has a non-vanishing spin-triplet gap from our results.
While the plus ($+$) symbols on the extrapolated curves, which mark
our corresponding estimates for the two QCPs at $\alpha^{c}_{1a}$ and
$\alpha^{c}_{1b}$ do not precisely lie on the $\Delta=0$ axis and mark
the points beyond which $\Delta$ becomes nonzero, the shapes of the
curves in this region clearly militates against using the results for
$\Delta$ to give accurate determinations of the critical values.
Nevertheless, our results for $\Delta$ are in good agreement with
those for $M$, and provide strong preliminary evidence that over the
entire intermediate range $\alpha^{c}_{1a} < \alpha < \alpha^{c}_{1b}$
the GS phase is gapped.

\begin{figure}
\begin{center}
\includegraphics[width=8.5cm]{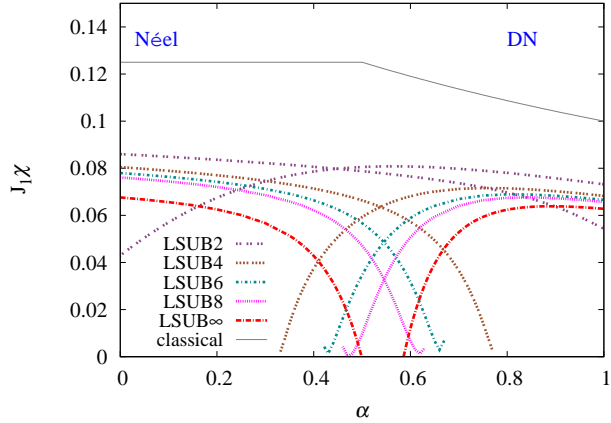}
\caption{CCM results for the zero-field transverse magnetic
  susceptibility $\chi$ (in units of $J_{1}^{-1}$, and where the gyromagnetic ratio $g\mu_{B}/\hbar=1$) as a function of the frustration parameter $\alpha \equiv
  J_{2}/J_{1}$, for the spin-$\frac{1}{2}$ $J_{1}$--$J_{2}$ model on a
  cross-striped square lattice, with $J_{1}>0$.  Results are shown based on both the canted quasiclassical N\'{e}el
  (left curves) and DN (right curves) AFM states as the CCM model
  states, in LSUB$m$ approximations with $m=2,4,6,8$.  Also shown are the extrapolation based on Eq.\ (\ref{Eq_X}) and the data set with
  $m=\{4,6,8\}$ as input and, for comparison, the corresponding classical result of Eq.\ (\ref{ExtMField_classical}).}
\label{M_Xcpty}
\end{center}
\end{figure}

In order both to corroborate these findings and to confirm the
critical values of the frustration parameter it is now convenient to
calculate the zero-energy (uniform) transverse magnetic susceptibility $\chi$.
Unlike for a system with magnetic LRO, a gapped system has no access
to the low-energy excitations that make the susceptibility to an
external magnetic field nonzero as the field becomes vanishingly
small.  Thus, at $T=0$, a gapped system has $\chi=0$
\cite{Mila:2000_M-Xcpty_spinGap,Bernu:2015_M-Xcpty_spinGap}.  In Fig.\
\ref{M_Xcpty} we show our LSUB$m$ results for $\chi$ based on both the
canted N\'{e}el and DN states as CCM model states.
As discussed previously, due to the reduced symmetries of those states
compared to their zero-field counterparts, we can only perform LSUB$m$
calculations now with $m \leq 8$.  Accordingly, in Fig.\
\ref{M_Xcpty}, we only now show a single LSUB$\infty$ extrapolation
based on the scheme of Eq.\ (\ref{Eq_X}) and the data set with
$m=\{4,6,8\}$ as input.

For both magnetically ordered phases the spin-$\frac{1}{2}$ system has
an appreciably lower value of $\chi$ for a given value of $\alpha$
than its classical ($s \rightarrow \infty$) counterpart, as may be
clearly seen from Fig.\ \ref{M_Xcpty}.  More interestingly, each of
the LSUB$m$ curves with $m > 2$ approaches zero very closely at an
upper (lower) critical value of $\alpha$ for the N\'e{e}l (DN) case.
Remarkably, in every case these critical values where $\chi(m) \rightarrow 0$ are
identical (to within $\pm 0.01$) to the corresponding LSUB$m$ values
where $M(m) \rightarrow 0$.  Unlike for the case of $\Delta$ where
there exists an appreciable unphysical regime where real solutions
still exist beyond the points where $M(m) \rightarrow 0$, in the case
of $\chi(m)$ the solutions terminate very rapidly after the points
where $\chi(m) \rightarrow 0$, with the values increasing to become
positive and nonzero again in these small regions.  This behaviour is
particularly clearly seen in Fig.\ \ref{M_Xcpty} for the two sets of
higher-order LSUB$m$ curves with $m=6,8$.

The above behaviour, particularly the very close agreement for the
critical values for $\alpha$ at which corresponding LSUB$m$ values of
$M(m)$ and $\chi(m)$ vanish, provides compelling evidence that the GS
phase with no magnetic LRO, lying between the N\'{e}el and DN phases,
is gapped everywhere.  Nevertheless, the QCPs obtained from the
LSUB$\infty$ curves for $\chi$ shown in Fig.\ \ref{M_Xcpty}, namely
$\alpha^{c}_{1a}=0.50$ and $\alpha^{c}_{1b}=0.59$, are slightly
different from our estimates $\alpha^{c}_{1a}=0.46(1)$ and
$\alpha^{c}_{1b}=0.615(5)$ obtained from $M$.  In the absence of a
comparable sensitivity analysis of our results for $\chi$, due to
being restricted to performing LSUB$m$ calculations with $m \leq 8$ in
this case, it is difficult to say with complete authority whether the
small discrepancies in the values obtained for the QCPs from $M$ and
$\chi$ are significant.  Given the very close agreements, however,
obtained at the individual LSUB$m$ levels with $m \leq 8$, it seems
most likely that the difference at the extrapolated levels is wholly
within the unknown errors of the extrapolation for $\chi$.
Nevertheless, we cannot entirely rule out the possibility that the
region intermediate between the magnetically ordered phases contains
both a gapped phase within an inner region determined from the
vanishing of $\chi$ on both sides and one or more gapless phases
outside this region but still inside the region determined from the
vanishing of the order parameter $M$ for the two quasiclassical
phases.  

\section{Conclusions}

We have found that the single critical point at
$\alpha^{{\rm cl}}=\frac{1}{2}$ in the classical $J_{1}$--$J_{2}$
model on the cross-striped square lattice, which separates two
different (viz., N\'{e}el and DN) states with AFM LRO, is split by
quantum fluctuations in the spin-$\frac{1}{2}$ version of the model
into two QCPs at $\alpha_{1a}^{c}=0.46(1)$ and
$\alpha_{1b}^{c}=0.615(5)$, at which the respective quasiclassical
forms of magnetic LRO vanish.  Further calculations of the triplet
spin gap and the zero-field transverse magnetic susceptibility have
provided convincing evidence that over most (and, most likely, all) of
the intermediate range $\alpha_{1a}^{c} < \alpha < \alpha_{1b}^{c}$
the stable GS phase is a gapped state without any discernible magnetic
LRO.  This finding is in complete agreement with earlier work
\cite{Bishop:2013_crossStripe} on the spin-$\frac{1}{2}$ model that
gave evidence of an intermediate state with plaquette valence-bond
crystalline order.

Since the sole approximation made in our CCM calculations is the
extrapolation of our series of LSUB$m$ approximants to the exact
$m \rightarrow \infty$ limit, and since we have been able to implement
the method here to very high orders (viz., with $m \leq 10$), our
results are inherently more accurate than can be attained in most
alternative treatments.  Furthermore, since the CCM, unlike most other
methods, explicitly preserves the Hellmann-Feynman theorem at all
levels of approximation, our results for the magnetic order parameter
$M$ and the zero-field transverse magnetic susceptibility $\chi$, for
example, are thus entirely compatible with each other at all LSUB$m$
levels.  This fact, together with our explicit LSUB$m$ sets of results
for $M$ and $\chi$, shown in Figs.\ \ref{M} and \ref{M_Xcpty}
respectively, strengthens our belief that over the entire range
$\alpha_{1a}^{c} < \alpha < \alpha_{1b}^{c}$ the stable GS phase is
gapped.  Nevertheless, even with these compelling arguments, we cannot
entirely rule out the presence of very narrow regimes near one or both
QCPs, $\alpha_{1a}^{c}$ and $\alpha_{1b}^{c}$, at which quasiclassical
AFM LRO vanishes, in which the GS phase is gapless.  However, if they
do exist, their extent in the parameter $\alpha$ is, undoubtedly,
limited to a small part of the intermediate regime.

\section*{ACKNOWLEDGMENT}
We thank the University of Minnesota Supercomputing
Institute for the grant of supercomputing facilities.  One of us (RFB) acknowledges the Leverhulme Trust (United Kingdom) for the award of an Emeritus Fellowship (EM-2015-007).

\section*{References}



\bibliographystyle{elsarticle-num} 
\bibliography{bib_general}


%\begin{thebibliography}{00}


%\bibitem{}

%\end{thebibliography}
\end{document}